\documentclass[aps,prx,showpacs,superscriptaddress,twocolumn,longbibliography]{revtex4}
\usepackage{amsfonts}
\usepackage{txfonts}
\usepackage{epsfig}

\def\be{\begin{equation}} \def\ee{\end{equation}}
\def\bea{\begin{eqnarray}} \def\eea{\end{eqnarray}}

\def\nn{\nonumber}

\begin{document}
\title{Simplified Topological Invariants for Interacting Insulators}

\author{Zhong Wang}

\affiliation{
Institute for Advanced Study, Tsinghua University, Beijing,  China, 100084}

\author{Shou-Cheng Zhang$^{1,}$}

\affiliation{
Department of Physics, Stanford University, CA 94305}

\date{\today}

\begin{abstract}

We propose general topological order parameters for interacting insulators in terms of the Green's function at zero frequency. They provide an unified description of various interacting topological insulators including the quantum anomalous Hall insulators and the time reversal invariant insulators in four, three and two dimensions. Since only Green's function at zero frequency is used, these topological order parameters can be evaluated efficiently by most numerical and analytical algorithms for strongly interacting systems.

\end{abstract}


\maketitle

\emph{Introduction.}
Topological insulators are new quantum states of matter whose characteristic property is the existence of both bulk energy gap and stable surface states\cite{qi2010a, moore2010, hasan2010,qi2011}. The stability of surface states are protected by the bulk topology, which in the non-interacting limit is described by Bloch band topological invariants such as the Thouless-Kohmoto- Nightingale-den Nijs(TKNN) invariant\cite{thouless1982} and the ${\rm Z}_2$ invariants\cite{kane2005b,moore2007,fu2006,fu2007b,Roy2009a,qi2008}. More recently, topological insulator with strong electron-electron interaction is becoming a central topic in the field\cite{raghu2008,shitade2009,zhang2009b,seradjeh2009,pesin2010,fidkowski2010,
li2010,dzero2010,rachel2010,zhang2011,neupert2012}. For general interacting systems, the topological order
parameters can be defined as the physical response function for the quantum Hall effect\cite{laughlin1983} and the topological magneto-electric effect\cite{qi2008}.
For actual evaluations of these physical response functions, it was proposed that Green's function is an useful tool in topological insulators\cite{wang2010b}, and there is much recent interest focused in this direction\cite{wang2011,wang2011a,gurarie2011,chen2011}. However, our original formula for
the topological order parameter\cite{wang2010b} is rather complicated; more recently, a much simpler formula was
obtained for the inversion symmetric interacting topological insulators\cite{wang2012}.

The main purpose of this paper is to obtain several simple and yet general topological order parameters for interacting topological insulators in an unified framework. They are expressed in terms of Green's function at zero frequency instead of the entire frequency domain. These invariants strongly resemble the conventional topological invariants such as the Chern number/TKNN invariant; yet, they are valid for
general interacting systems. Current proposals for the quantum anomalous Hall (QAH) insulators\cite{qi2005,liu2008a,yu2010a} require magnetic
order, which is only possible for interacting
systems. Our proposed topological order parameter can greatly help the search for realistic materials. Among our central results are Eq.(\ref{chern}), Eq.(\ref{derivation}), Eq.(\ref{2ndchern}), Eq.(\ref{p3}), and Eq.(\ref{pfaffian}), all of which are expressed in terms of Green's function at $i\omega=0$. In most numerical
algorithms of strongly interacting systems, it is much easier to obtain the Green's function at zero frequency rather
than all frequencies. Therefore, our new formulas present a significant improvement over the previous result\cite{wang2010b}. We would also like to point out that the formulas given in this paper is not directly applicable to fractional topological insulators with nontrivial ground states degeneracy, which will be left to future studies.

\emph{Topological order parameter for interacting QAH insulators.}
The conventional topological invariant for two-dimensional(2d) non-interacting quantum (anomalous) Hall states (or ``Chern insulator'') is the TKNN invariant\cite{thouless1982}, which is also called the first Chern number in mathematic literatue. Explicitly, it is an integral over the momentum space (namely the first Brillouin zone)
\bea  c_1 = \frac{1}{2\pi} \int d^2k f_{xy} \label{tknn} \eea where $f_{ij}=\partial_i a_j - \partial_j a_i$, and $a_i =-i\sum_\alpha \langle \psi^\alpha(k)|\partial_{k_i}|\psi^\alpha(k)\rangle$, where $\alpha$ runs through all the occupied bands.
However, because its fundamentally dependence on the Bloch state $|\psi^\alpha(k)\rangle$, it applies only to non-interacting systems. There is an interesting generalization to interacting systems using twisted boundary condition\cite{niu1985}, which is nonetheless
difficult to compute and is not easy to generalize to ${\rm Z}_2$ insulators.   There is another integer topological invariant expressed in term of Green's function rather than the Bloch state\cite{ishikawa1986,wang2010b,volovik2003}  \bea N_2=
 \frac{1}{24\pi^2} \int dk_0 d^2k {\rm Tr}[\epsilon^{\mu
\nu \rho} G\partial_\mu G^{-1}G\partial_\nu G^{-1} G\partial_\rho
G^{-1}] \label{n2} \eea where $\mu,\nu,\rho$ run through $k_0,k_1,k_2$, with $k_0=i\omega$ referring to the Matsubara frequency (imaginary frequency).
Throughout this paper, the Green's functions are the Matsubara Green's functions, though our final formulas are also applicable using real frequency. For instance, a non-interacting system with Hamiltonian $H= \sum_k c^\dag_k h(k) c_k$ has $G(i\omega,k)=1/(i\omega-h(k))$, where $h(k)$ is generally a $k$-dependent matrix, and $c_k$ is a column vector of fermion operators. The discrete Matsubara frequency becomes continuous in the zero temperature limit which we take.

Eq.(\ref{n2}) has the severe disadvantage that it involves a frequency integral.
In most numerical algorithms, it is very difficult to obtain the dynamic Green's function at all frequencies as
required by Eq.(\ref{n2}). Now we shall show that it is possible, without any approximations, to evaluate
Eq.(\ref{n2}) with only the Green's function at zero frequency. Our new formula is much easier for practical
calculations, and is accessible for most numerical algorithms.

Let us start from the formalism presented in our previous work\cite{wang2012}. We diagonalize the inverse Green's function as
\bea G^{-1}(i\omega,k)|\alpha(i\omega,k)\rangle &=& \mu_\alpha (i\omega,k)|\alpha(i\omega,k)\rangle  \label{eigin} \eea The eigenvectors of $G$ are the same as those of $G^{-1}$, with eigenvalues $\mu_\alpha^{-1}$, therefore, we can also formulate our approach by diagonalizing $G$ instead of $G^{-1}$.
From the Lehmann representation, we can show that Green's function satisfies the equation
\bea (G^{-1})^\dag(i\omega,k)=G^{-1}(-i\omega,k) \label{hermitian} \eea
from which it follows that
\bea (G^{-1})^\dag(0,k)=G^{-1}(0,k) \eea
Therefore, $\mu_\alpha(0,k)$ are real numbers. The eigenvectors $|\alpha(0,k)\rangle$ can be divided into two types: those with $\mu_\alpha(0,k)>0$ are called ``Right-zero (R-zero)'', while those with $\mu_\alpha(0,k)<0$ are called ``Left-zero (L-zero)''.  All the R-zeros span a subspace at each $k$, which we call the ``R-space''. Similarly we can define the ``L-space''.  Because the eigenvectors corresponding to different eigenvalues  of a Hermitian matrix are orthogonal, the R-space has the crucial property that all the vectors within it are orthogonal to those within the L-space. Therefore, the Chern number/TKNN number can be defined for the R-space in the conventional way.  More explicitly, we present one of the central result of this paper, which we shall call the ``generalized TKNN invariant'', or the ``generalized Chern number''
\bea  C_1 = \frac{1}{2\pi} \int d^2k \mathcal{F}_{xy} \label{chern} \eea where $\mathcal{F}_{ij}=\partial_i \mathcal{A}_j - \partial_j \mathcal{A}_i$, and $\mathcal{A}_i = -i\sum_{\alpha\in R-space}\langle k\alpha |\partial_{k_i}|k\alpha\rangle$.  Here $|k\alpha\rangle$ is an orthonormal basis of the R-space.  The simplest basis choice is $|k\alpha\rangle=|\alpha(i\omega=0,k)\rangle$(i.e. normalized R-zero).   Throughout this paper, the lower case expressions $c_1,c_2,a_i,f_{ij}$ are reserved for the non-interacting Chern numbers, Berry connection and curvature, while the upper case expressions $C_1,C_2,\mathcal{A}_i,\mathcal{F}_{ij}$ refer to the generalized Chern numbers, generalized Berry connection and curvature for generally interacting systems.

The expression in Eq.(\ref{chern}) reduces to the conventional TKNN invariant in the non-interacting limit. However, it is essentially different in that it is defined in terms of R-zeros and R-space of the Green's function of a many-body system rather than the Bloch states of a non-interacting Hamiltonian.   The definition of R-zero and R-space is valid in the presence of general electron-electron interaction, and thus this topological order parameter is a highly nontrivial generalization of the TKNN invariant to the interacting insulators, maintaining the elegant form of Chern number.
Mathematically, Eq.(\ref{chern}) is the first Chern number of an $U(N)$ fiber bundle, whose bundle at each $k$ is exactly the R-space. Therefore, this quantity is a topologically invariant.

One can also define an ``effective Hamiltonian'' $h_{eff}(k)\equiv -G^{-1}(0,k)$, and define R-zeros as eigenvectors of this ``effective Hamiltonian''. This is just another language to formulate the same result, which we shall use later
in the derivation of our formula.

The characteristic physical observable of the 2d correlated QAH insulators is the Hall conductance \bea \sigma_{xy} = C_1\frac{e^2}{h} \eea which can be measured in experiments. As we shall show in the next section, the coefficient $C_1$ here is exactly the one in Eq.(\ref{chern}). The simple formula Eq.(\ref{chern}) can be easily applied to
the interacting quantum (anomalous) Hall systems with integer quantum Hall effect. However, it should be mentioned\cite{wang2010b} that this description cannot be directly applied to the fractional quantum Hall states which have nontrivial ground state degeneracy. The same limitation is also true for the frequency integral formula in Eq.(\ref{n2}). We will leave the possibility of extending our approach to cases with ground state degeneracy for future studies.

\emph{Derivation of the formula.}
Eq.(\ref{chern}) is itself a topological invariant that can be applied to interacting insulators in 2d that break time reversal symmetry. In this section, we would like to show that it is indeed the quantum Hall conductance, namely,
$N_2=C_1$, in other words,  Eq.(\ref{chern}) can be derived from Eq.(\ref{n2}).
Let us begin with the Lehmann representation of Matsubara Green's function (in the zero temperature limit) \bea G_{\alpha\beta}(i\omega,k) = \sum_m[\frac{\langle 0|c_{k\alpha}|m\rangle \langle m|c_{k\beta}^\dag |0\rangle }   { i\omega -(E_m-E_0)} +  \frac{ \langle m|c_{k\alpha}|0\rangle \langle 0|c_{k\beta}^\dag |m\rangle } {i\omega +(E_m-E_0)}]\eea where $|m\rangle$ are the exact eigenvectors of $K=H-\mu N$ ($H$ is the many-body Hamiltonian, and $\mu$ and $N$ is the chemical potential and fermion number respectively) with eigenvalues $E_m$, and $|0\rangle$ is the ground state. We can make the summation over $|m\rangle$ well defined by putting the system on a large but finite two dimensional torus so that the eigenvalues are discrete, otherwise the summation is replaced by integral.  Note that we have assumed that there is only a single ground state.  For our purpose, we decompose $G(i\omega,k)=G_1+iG_2$ with both $G_1$ and $G_2$ Hermitian. Explicitly, we have \bea (G_2)_{\alpha\beta} &=& -\sum_m\frac{\omega}{\omega^2+(E_m-E_0)^2} \nonumber \\ && \times[\langle 0|c_{k\alpha}|m\rangle \langle m|c_{k\beta}^\dag |0\rangle +  \langle m|c_{k\alpha}|0\rangle \langle 0|c_{k\beta}^\dag |m\rangle ] \nonumber \\ &=& -\sum_m d_m [u_{m\alpha}^* (k) u_{m\beta}(k) + v_{m\alpha}^*(k) v_{m\beta}(k)]\eea
where we have defined $u_{m\alpha}(k) =\langle m|c_{k\alpha}^\dag|0\rangle$, $v_{m\alpha} =\langle 0|c_{k\alpha}^\dag|m\rangle$ and $d_m(i\omega)=\omega /[\omega^2+(E_m-E_0)^2]$ to simply the expressions. It is easy to see that ${\rm sign}(d_m)= {\rm sign}(\omega)$.
Now we can calculate the expectation of $G_2$ with an arbitrary vector $|a\rangle$ as $\langle a|G_2|a\rangle=\sum_{\alpha\beta} a^*_\alpha (G_2)_{\alpha\beta} a_\beta = -\sum_m d_m[|\sum_\alpha a_\alpha u_{m\alpha}|^2 + |\sum_\alpha a_\alpha v_{m\alpha}|^2]$.  From this we can see \bea {\rm sign}(\langle a|G_2(i\omega,k)|a\rangle)=-{\rm sign}(\omega) \label{sign} \eea
As a side remark, if $|a\rangle$ is an eigenvector of $G=G_1+iG_2$ with eigenvalue $\mu_a^{-1}$, then we have $\mu_a^{-1}=\langle a|a\rangle^{-1}\langle a|(G_1+iG_2)|a\rangle$. Due to the fact that $G_1$ and $G_2$ are Hermitian, we have ${\rm Im}(\mu_a^{-1}) = \langle a|a\rangle^{-1}\langle a|G_2|a\rangle $, thus it follows from Eq.(\ref{sign}) that
${\rm sign}[{\rm Im}(\mu_a^{-1} (i\omega))]= -{\rm sign} (\omega)$.  It follows finally that \bea {\rm sign}[{\rm Im}(\mu_a (i\omega,k))]= {\rm sign}(\omega) \label{sign-1}\eea
With these preparations, we are approaching the central part of our calculation. The key idea is to
introduce a smooth deformation of $G(i\omega,k)$ parameterized by $\lambda\in[0,1]$ as follows  \bea G(i\omega,k,\lambda) = (1-\lambda)G(i\omega,k) +\lambda [i\omega+G^{-1}(0,k)]^{-1} \label{deform} \eea We now show that this deformation does not contain singularity, or equivalently, we must show that all eigenvalues of $G(i\omega,k,\lambda)$ are nonzero. This can be seen as follows. First, when $i\omega=0$, we have $G(0,k,\lambda)=G(0,k)$, whose eigenvalues are nonzero by our assumption that the Green's function $G(i\omega,k)$ is nonsingular. Second, we consider $i\omega\neq 0$. Suppose that \bea G(i\omega,k,\lambda)|\alpha(i\omega,k,\lambda)\rangle=\mu_\alpha^{-1} (i\omega,k,\lambda)|\alpha(i\omega,k,\lambda)\rangle \nn \eea then we have \bea \mu_\alpha^{-1}(i\omega,k,\lambda) = \langle \alpha |\alpha \rangle^{-1} \langle \alpha  |G(i\omega,k,\lambda) |\alpha \rangle \nn \eea
The imaginary part of this equation can be written down as
\bea {\rm Im}[\mu_\alpha^{-1}(i\omega,k,\lambda)]= && \langle \alpha|\alpha\rangle^{-1}[(1-\lambda)\langle \alpha|G_2(i\omega,k)|\alpha\rangle \nn \\ &&-\lambda \omega\sum_s|\alpha_s|^2(\omega^2+\epsilon_s^2)^{-1}] \nn \eea
where we have expanded $|\alpha(i\omega,k,\lambda)\rangle=\sum_s \alpha_s(i\omega,k,\lambda) |s(k)\rangle$, in which $|s(k)\rangle$ are orthonormal eigenvectors of $-G^{-1}(0,k)$ with eigenvalues $\epsilon_s(k)$. It is easy to see that ${\rm Im}[\mu_\alpha^{-1}(i\omega,k,\lambda)]$ is always nonzero thanks to Eq.(\ref{sign}).
Summarizing the above calculation, we can see that all eigenvalues of $G(i\omega,k,\lambda)$ are nonzero, therefore the deformation in Eq.(\ref{deform}) is smooth.  Note that throughout this calculation we consider the imaginary frequency Green's function, otherwise Green's function cannot be so well-behaved.  A geometrical visualization of the deformation in Eq.(\ref{deform}) can be given as follows. Because of Eq.(\ref{sign-1}), the $\mu_\alpha(i\omega) (\omega\in(-\infty,+\infty))$ curves\cite{wang2012} on the complex plane do not cross the real axis when $i\omega\neq 0$, therefore we can smoothly deform them to straight lines parallel with the imaginary axis, keeping the R/L-zero unchanged in the deformation. This leads exactly to $G(i\omega,k,\lambda=1)$.

Because $N_2$ is a topological invariant, namely that it is unchanged under smooth deformations of $G$, we have $N_2(\lambda=0)=N_2(\lambda=1)$.  Therefore, to calculate $N_2=N_2(\lambda=0)$, we just need to calculate $N_2(\lambda=1)$, which is equivalent to the calculation for an effective noninteracting system with $h_{eff}(k)=-G^{-1}(0,k)$. It is a straightforward calculation to obtain  $N_2(\lambda=1)=C_1$.
This completes the derivation of \bea N_2=C_1 \label{derivation}\eea which is a precise identity between Eq.(\ref{n2}) and Eq.(\ref{chern}).

\emph{Four-dimensional topological insulators.}
The 2d physics discussed above can be generalized to 4d. In 4d, there is time reversal invariant topological insulator classified by integer ${\rm Z}$. The continuous model for such topological insulators was first proposed in Ref.\cite{zhang2001}, while lattices models can be found in Ref.\cite{qi2008}. For non-interacting insulators in 4d, the natural topological invariant is the second Chern number in the momentum space expressed as\cite{qi2008}
\bea
c_2&=&\frac1{32\pi^2}\int d^4k\epsilon^{ijkl}{\rm
tr}\left[f_{ij}f_{kl}\right]\label{2ndtknn} \eea
with
\bea && f^{\alpha\beta}_{ij}=\partial_i
a^{\alpha\beta}_j-\partial_j
a^{\alpha\beta}_i+i\left[a_i,a_j\right]^{\alpha\beta},\nonumber\\
&&  a_i^{\alpha\beta}(k) = -i\left\langle \psi^\alpha(k)\right|\frac{\partial}{\partial k_i }\left|\psi^\beta(k)\right\rangle\nonumber
\eea
where $i,j,k,l=1,2,3,4.$ The index $\alpha$ in $a_i^{\alpha\beta}$ refers to the occupied bands
of the Bloch states $|\psi^\alpha(k)\rangle$. The Berry connection
$a_i^{\alpha\beta}$ is a non-abelian gauge field potential, and
$f^{\alpha\beta}_{ij}$ is the associated non-abelian field
strength.
Analogous to the 2d case, for 4d interacting insulators there is a topological invariant analogous to Eq.(\ref{n2}), expressed in terms of the interacting Green's function\cite{wang2010b,qi2008}
\bea N_4 &\equiv&
\frac{1}{480\pi^3} \int d^{5}k
\textrm{Tr}[\epsilon^{\mu \nu \rho \sigma \tau}
G\partial_{\mu}G^{-1} G\partial_{\nu}G^{-1} G\partial_{\rho}G^{-1}
\nn \\ &\,& \times G\partial_{\sigma}G^{-1} G\partial_{\tau}G^{-1}]
\,   \label{n4} \eea
This topological order parameter directly measures the generalized quantum
Hall effect in 4d\cite{zhang2001,qi2008}, and it is related to the homotopy group $\pi_5({\rm GL}(N,{\rm C}))= \mathbb{Z}$\cite{wang2010b} for sufficiently large $N$. The difficulty with Eq.(\ref{n4}) is again the frequency integral over $(-i\infty,+i\infty)$. This problem can be solved in the same way as its 2d analog. We are thus led to another central result of this paper, namely the
topological order parameter for an interacting Chern insulator in 4d, expressed as
\bea
C_2&=&\frac1{32\pi^2}\int d^4k\epsilon^{ijkl}{\rm
tr}\left[\mathcal{F}_{ij}\mathcal{F}_{kl}\right]\label{2ndchern} \eea
with
\bea  && \mathcal{F}^{\alpha\beta}_{ij}=\partial_i
\mathcal{A}^{\alpha\beta}_j-\partial_j
\mathcal{A}^{\alpha\beta}_i+i\left[\mathcal{A}_i,\mathcal{A}_j\right]^{\alpha\beta},\nonumber\\ &&
\mathcal{A}_i^{\alpha\beta}(k) = -i\langle k \alpha|\frac{\partial}{\partial k_i }|k\beta\rangle\nonumber
\eea
where $|k\alpha\rangle$ is the same as that defined below Eq.(\ref{chern}), namely that $|k\alpha\rangle$ is an orthonormal basis of the R-space spanned by R-zeros.  The derivation of $N_4=C_2$ is a straightforward generalization of that of Eq.(\ref{derivation}), which we shall not repeat here.

\emph{${\rm Z}_2$ topological invariants for interacting insulators in three  and two spatial dimensions.}
It was proposed in Ref.\cite{wang2010b} that a natural topological invariant for 3d ${\rm Z}_2$ insulator is the
topological magneto-electric coefficient\cite{wang2010b,wang2012},
\bea 2P_{3} &=& W(G)|_{R\times T^4} \nonumber \\
&\equiv & \frac{1}{480\pi^3} \int_{-\pi}^{\pi}d^5k \textrm{Tr} [ \epsilon^{\mu \nu \rho \sigma \tau}
G\partial_{\mu}G^{-1}  \nonumber
\\  &\,& \times  G\partial_{\nu}G^{-1} G\partial_{\rho}G^{-1} G\partial_{\sigma}G^{-1}
G\partial_{\tau}G^{-1}] \label{wzw} \eea
where the integer $W(G)|_{R\times T^4}$ is the ``winding number'' of the mapping from frequency-momenta space $R\times T^4$ to $GL(N,C)$, $k_0=i\omega$ is the imaginary frequency, and $k_4$ is the dimensional extension
parameter similar to the Wess-Zumino-Witten parameter in non-linear $\sigma$ models. The reference function $G(k_0,k_1,k_2,k_3,\pi)$ is trivially diagonal\cite{wang2010b}.  Due to the ambiguity of the
dimensional extension, the integer in Eq.(\ref{wzw}) reduces to ${\rm Z}_2$ equivalent classes\cite{wang2010b}.
For insulators with inversion symmetry, Eq.(\ref{wzw}) is further simplified to a product of parity of R-zeros\cite{wang2012}. This major simplification enables practical numerical calculations using Green's function, see e.g. \cite{go2012}.

Eq.(\ref{wzw}), although an elegant ${\rm Z}_2$ invariant, is unsatisfactory because the needs of the dimensional extension and the frequency integral. Now we shall obtain ${\rm Z}_2$ topological invariants without these two disadvantages. This is outlined as follows. Let us start from Eq.(\ref{wzw}) and dimensionally extend $G(k_0,k_1,k_2,k_3)$ to $G(k_0,k_1,k_2,k_3,k_4)$, where $k_4\in [-\pi,\pi]$ is the WZW like dimensional extension parameter. Zero frequency function $G(0,k_1,k_2,k_3,k_4)$ are chosen to be Hermitian, therefore, R-zeros can be extended to the extended momenta space $T^4$.  By a calculation analogous to the the derivation of $N_4=C_2$, Eq.(\ref{wzw}) can be simplified into an interacting topological order parameter in 3d as
\begin{eqnarray} P_{3} = C_2/2=\frac{1}{32\pi^2}\int_0^{\pi}dk_4\int_{-\pi}^{\pi} d^3k\epsilon^{ijkl}{\rm
tr}[\mathcal{F}_{ij}\mathcal{F}_{kl}] \nonumber \\
= \frac{1}{8\pi^{2}} \int
d^{3}k\epsilon^{ijk} \textrm{Tr}\{[\partial_i\mathcal{A}_{j}(k)+\frac{2}{3}i  \mathcal{A}_{i}(k)\mathcal{A}_{j}(k))]\mathcal{A}_{k}(k)\} \label{p3}
\end{eqnarray}
Here the Berry connection is defined in terms of zero frequency Green's function in the same way as Eq.(\ref{chern}) and Eq.(\ref{2ndchern}).
Eq.(\ref{p3}) generalizes the formula first obtained by Qi, Hughes and Zhang for the non-interacting
system\cite{qi2008}.
Because the Green's function instead of the Bloch states are used,
Eq.(\ref{p3}) is defined for interacting topological insulators in 3d.  Eq.(\ref{p3}) has great advantage compared to Eq.(\ref{wzw}) because neither dimensional extension nor frequency integral is needed. In the non-interacting limit, the R-zeros become Bloch states, and  Eq.(\ref{p3}) is thus reduced to the Chern-Simons(CS) invariant defined in Ref.{\cite{qi2008}.  As was shown in Ref.\cite{wang2010a}, the non-interacting CS invariant\cite{qi2008} is equivalent to the Fu-Kane's Pfaffian invariant\cite{fu2006}, it is thus natural to obtain an Pfaffian invariant $\Delta$ for interacting topological insulators, which is expressed in terms of R-zeros as \bea (-1)^\Delta=\prod_{\Gamma_i={\rm TRIM}} \frac{  \sqrt{{\rm det}B(\Gamma_i)} } {{\rm Pf}(B(\Gamma_i))} \label{pfaffian} \eea
Here, ``TRIM'' refers to time reversal invariant momenta, the matrix $B(k)$ is defined by $B_{\alpha\beta}=\langle -k\alpha|\hat{T}|k\beta\rangle$, where $\hat{T}$ is the time reversal operation, and $|k\alpha\rangle$ is an orthonormal basis of R-space (the simplest choice is $|k\alpha\rangle=|\alpha(i\omega=0,k)\rangle$).  Eq.(\ref{p3}) and Eq.(\ref{pfaffian}), which are ${\rm Z}_2$ invariants defined for interacting insulators, are among the central results of this paper. They have the great advantage that only zero frequency Green's function is needed. The ambiguity of the square root in Eq.(\ref{pfaffian}) can be avoided if one rewrite it as an integral following Ref.\cite{fu2006}. We also mention that there are other formulas equivalent to Eq.(\ref{pfaffian}), e.g. number of zeros of Pfaffian in the half Brillouin zone, which parallel the non-interacting formulas, with Bloch states replaced by vectors in R-space.
The derivation from Eq.(\ref{p3}) to Eq.(\ref{pfaffian}) follows exactly the calculations in Ref.\cite{wang2010a}, again with the Bloch states replaced by vectors in the R-space. A similar expression as Eq.(\ref{pfaffian}) can also be defined for 2d interacting topological insulators.  For insulators with inversion symmetry, starting from Eq.(\ref{p3}) or Eq.(\ref{pfaffian}) one can derive the parity formula in Ref.\cite{wang2012}, which was originally derived from  Eq.(\ref{wzw}) directly. The parity formula\cite{wang2012} is most convenient for practical calculation if the insulator has inversion symmetry. However, if the insulator has no inversion symmetry, we need to use the more general formulas in Eq.(\ref{p3}) or Eq.(\ref{pfaffian}) proposed in the present paper.

\emph{Conclusion.} In this work we present a general framework to describe interacting insulators in terms of the Green's function at zero frequency. Our central results include the
topological order parameters for the quantum anomalous Hall insulator, interacting Landau level
systems with integer quantum Hall effect, and time reversal invariant interacting topological insulators in
4d, 3d and 2d. These formulas greatly simplify numerical and analytical calculations.

We would like to thank Xiaoliang Qi for collaborations in earlier works which paved the way for the present
simplifications, and Xi Dai for invaluable discussion on the ${\rm Z}_2$ topological invariant. ZW is supported by Tsinghua University (No. 553401001). SCZ is supported by the NSF under grant numbers DMR-0904264 and the Keck Foundation.

\bibliography{TKNN}

\end{document}